\documentclass[twocolumn, showpacs, prl]{revtex4}                             
                                                                              
\usepackage{epsfig}                                                           
\usepackage{amssymb}                                                          
                                                                              
\begin{document}                                                              
                                                                              
\title{Phase Field Modeling of Fast Crack Propagation}                        
\author{Robert Spatschek}                                                     
\email{r.spatschek@fz-juelich.de}                                             
\author{Miks Hartmann}                                                        
\author{Efim Brener}                                                          
\author{Heiner M\"uller-Krumbhaar}                                            
\affiliation{Institut f\"ur Festk\"orperforschung. Forschungszentrum          
J\"ulich, D-52425 J\"ulich, Germany}                                          
\author{Klaus Kassner}                                                        
\affiliation{Institut f\"ur Theoretische Physik, Universit\"at Magdeburg,     
D-39016 Magdeburg, Germany}                                                   
\date{\today{}}                                                               
                                                                              
\begin{abstract}                                                              
We present a continuum theory which predicts the steady state propagation of  
cracks.                                                                       
The theory overcomes the usual problem of a finite time cusp singularity of   
the Grinfeld instability by the inclusion of elastodynamic effects which      
restore selection of the steady state tip radius and velocity.                
We developed a phase field model for elastically induced phase transitions;    
in the limit of small or vanishing elastic coefficients in the new phase, 
fracture can be studied. 
The simulations confirm analytical predictions for fast crack             
propagation.                                                                  
\end{abstract}                                                                
                                                                              
\pacs{62.20.Mk, 46.50.+a, 46.15.-x, 47.54.+r}                                 
                                                                              
\maketitle                                                                    
                                                                              
Understanding the day-to-day phenomenon of fracture is a major challenge for  
solid state physics and materials science.                                    
Starting with the early idea of Griffith \cite{Griffith21}, who realized that 
crack growth is a competition between a release of elastic energy and an      
increase of surface energy, various approaches have been developed to         
describe the striking features of cracks \cite{Freund98}.                     
Usually, the motion of cracks is understood on the level of breaking bonds at 
sharp tips, and obviously theoretical predictions depend sensitively on the   
underlying empirical models of the atomic properties (see for example         
\cite{Hauch99}).                                                              
Plastic effects, however, lead to extended crack tips (finite tip radius      
$r_0$), and it is conceivable that for example fracture in gels can be        
described macroscopically.                                                    
Then a full modeling of fracture should not only determine the crack speed    
but also the crack shape self-consistently.                                   
                                                                              
Recent phase field models go beyond the microscopic limit of discrete models with
broken rotational symmetry, and encompass much of the expected    
behavior of cracks \cite{Karma,Henry04};                                      
these models are close in spirit though different in details    
with respect to earlier approaches \cite{Aranson00}.  
However, the scale of the appearing patterns is always
dictated by the phase field interface width, and thus these models
have problems in the sharp interface limit. 
Other descriptions are based on macroscopic equations of motion but suffer    
from inherent finite time singularities which do not allow steady state crack 
growth unless the tip radius is limited by the phase field interface width    
\cite{Kassner01}.                                                             
Numerical approaches which are not based on a phase field provide 
a selection mechanism by the introduction of complicated nonlinear terms in   
the elastic energy for high strains in the tip region \cite{Marconi05},
requiring additional parameters.       
                                                                              
It is therefore highly desirable to look for minimal models 
of fracture which are free from microscopic details and which are based
on well established thermodynamical concepts. 
This is also motivated by experimental results showing that many       
features of crack growth are rather generic \cite{Fineberg99};                
among them is the saturation of the steady state velocity appreciably below   
the Rayleigh speed and a tip splitting for high applied tension.              
                                                                              
Already in our previous publication \cite{Brener03} we emphasized a           
connection between fracture mechanics and elastically induced surface         
diffusion processes:                                                          
the Asaro-Tiller-Grinfeld (ATG) instability \cite{ATG} appears to be
a good starting point   
for the quest for a ``macroscopic'' theory of fracture, where  
crack growth is mediated by surface diffusion along the crack surfaces.    
Plasticity effects are contained in this description in a            
lubrication approximation, provided that the region of inelasticity is        
relatively thin in comparison to the tip radius.                              
                                                                              
Here we propose a similar approach which describes crack growth by a phase    
transition model. In fact, it produces a non-trivial dynamical selection
of the radius of the crack tip.                                                             
Instead of cracks filled with vacuum, we consider for the moment 
a soft condensed phase inside   
the crack which is growing at the expense of a brittle material               
\cite{Kassner01}.                                                             
                                                                              
The difference in the chemical potentials between two phases at an      
interface is \cite{Nozieres93}                                                
\begin{equation} \label{intro:eq3}                                            
\Delta\mu = \Omega \left( \frac{1}{2} \sigma_{jk}u_{jk} - \gamma\kappa        
\right),                                                                      
\end{equation}                                                                
provided that the soft phase is stress free because of negligible elastic     
moduli.                                                                       
Then the surface of the crack is free of normal and shear stresses.           
We assume for simplicity the mass density $\rho$ to be  equal in both phases  
and the elastic displacements to be continuous at the interface, which means  
that the two phases are coherent.                                             
Also, we assume a two-dimensional geometry.                                   
The interfacial energy per unit area is $\gamma$, and the interface curvature 
$\kappa$ is positive if the crack shape is convex.                            
$\Omega$ is the atomic volume, $\sigma_{jk}$ and $u_{ik}$ stress and strain   
tensor respectively.                                                          
Stress and strain are connected by Hooke's law for isotropic elasticity,      
$\sigma_{kj} = 2\mu u_{kj} + \lambda u_{ll} \delta_{kj}$,                     
with the Lam\'e coefficient $\lambda$ and the shear modulus $\mu$.                 
Alternatively, we use Young's modulus $E=\mu(3\lambda+2\mu)/(\lambda+\mu)$    
and the Poisson ratio $\nu=\lambda/2(\lambda+\mu)$ as elastic constants.      
                                                                              
For phase transitions, the motion of the interface is locally expressed by    
the normal velocity                                                           
\begin{equation} \label{intro:eq4}                                            
v_n = \frac{D}{\gamma\Omega} \Delta\mu                                        
\end{equation}                                                                
with a kinetic coefficient $D$ with dimension $[D] = {\rm m}^2 {\rm s}^{-1}$. 
                                                                              
It is known that nonhydrostatic stresses $P$ at a nominally flat interface    
lead to the ATG instability:                                                  
Long-wave perturbations of a flat interface diminish the total energy of the  
system, whereas short-wave perturbations are hampered by surface energy.      
The characteristic length scale of this instability, $L_G\sim                 
E\gamma/P^2(1-\nu^2)$, is identical to the        
Griffith length of a crack, up to a numerical prefactor.    
This instability leads to a finite time cusp singularity in the framework of  
the static theory of elasticity \cite{Yang}:                                  
The tip radius decreases to zero and simultaneously the tip velocity grows    
indefinitely.                                                                 
In this sense, the advancing notches can be interpreted as cracks.            
We explained the unphysical breakdown of the late stage of the ATG            
instability in our previous publication for surface diffusion                 
\cite{Brener03}:                                                              
Here, similar arguments apply, and detailed counting arguments will be        
derived below proving that simultaneous selection of the tip radius and       
velocity is impossible in the framework of the static theory of elasticity.   
                                                                              
\begin{figure}                                                                
\begin{center}                                                                
\epsfig{file=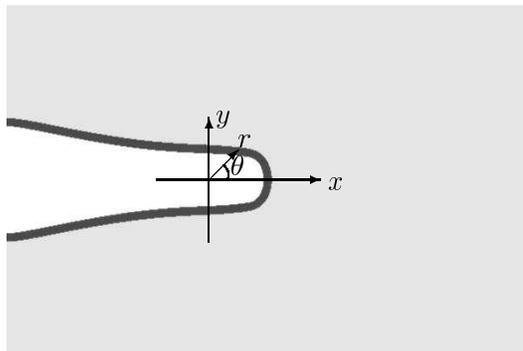, width=7cm}
\caption{Steady state growth of a crack in a strip, as obtained from phase-field
simulations. The grey color  
corresponds to the solid phase. A constant displacement is prescribed at the  
upper and lower boundary of the system. The parameters used here are          
$\Delta=2.03$ and $D/\epsilon v_R=6.18$; the system size is $600\times 400$.  
The resulting velocity is $v/v_R = 0.71$ and the radius $r_0=0.69             
D/v_R=4.26\epsilon$ are dynamically selected.}                                                         
\label{fig2}                                                                  
\end{center}                                                                  
\end{figure}

The elastic problem of a semi-infinite mathematical cut in an infinite (two   
dimensional) solid is exactly solved by a square root singularity of          
stresses, $\sigma_{ij}=K f_{0,ij}(\theta)/(2\pi r)^{1/2}$, using polar         
coordinates as depicted in Fig.\ \ref{fig2}.                                  
Here $K$ is the stress intensity factor (static or dynamic), $f_{0,ij}$ the   
universal angular distribution depending only on the mode of loading (for     
brevity, we suppress the velocity dependence $v/v_R$; $v_R$ is the Rayleigh   
speed);                                                                       
we concentrate on cracks of ''mode I''-type here.  
For an extended crack in an infinite environment, the square root singularity 
is only the slowest decaying mode;                                            
other powers can be present, and the total field can be interpreted as an     
expansion in the set of eigenfunctions of a straight crack:                   
\begin{equation} \label{neartip:eq1}                                          
\sigma_{ij}(r, \theta)=\frac{K}{(2\pi r)^{1/2}} \sum_{n=0}^{\infty} \frac{A_n 
f_{n,ij}(\theta)}{r^n}.                                                       
\end{equation}                                                                
Far field conditions imply $A_0=1$, whereas the other coefficients $A_n$ are  
determined by the boundary conditions of vanishing shear and normal stress on 
the crack;                                                                    
modes with ascending powers of $r$ are forbidden by boundary conditions, and  
even a homogeneous stress $P$ cannot be present in an infinite system         
(divergent strip width $L$), because this would lead to a diverging stress    
intensity factor $K\sim PL^{1/2}$.                                            
All higher order corrections in Eq.\ (\ref{neartip:eq1}) scale as $A_n\sim    
r_0^n$ and vanish in the sharp tip limit.                                     
From this equation follows readily that the stresses on the crack surface    
scale as $\sigma\sim r_0^{-1/2}$.                                             
                                                                              
We consider a crack as it might have developed in the late stage of the ATG   
instability, as depicted in Fig.\ \ref{fig2}.                                 
The macroscopic length of the crack is not considered here, and instead the   
stress intensity factor $K$ is given.                                         
At first, we demonstrate that steady state growth using only the {\em static} 
theory of elasticity is impossible;                                           
in fact, this is the reason for the mentioned cusp singularity.               
                                                                              
Assume that $y(x)$ in Cartesian or $r(\theta)$ in polar coordinates describes 
the steady state shape of a crack in the co-moving frame of reference,        
corresponding to a specific tip radius $r_0$ and velocity $v$.                
According to the results above, both contributions to the chemical potential  
Eq.\ (\ref{intro:eq3}) scale as $\mu\sim 1/r_0$ and thus by virtue of the     
equation of motion (\ref{intro:eq4}), $v\sim 1/r_0$.                          
Hence a rescaling of the steady state equation is possible:                   
Increasing the crack size by a factor $\alpha$ simply reduces the steady      
state velocity by $1/\alpha$ and vice versa.                                  
In other words, the explicit length scale $r_0$ drops out of the equations,   
and only $vr_0/D$ remains as dimensionless parameter.                         
All other parameters combine to the dimensionless driving force               
$\Delta=K^2(1-\nu^2)/2E\gamma$;                                               
$\Delta=1$ corresponds to the Griffith point.                                 
Notice that this rescaling is only possible in the framework of the static    
theory of elasticity;                                                         
otherwise, the stress field itself becomes velocity dependent, introducing    
the ratio $v/v_R$ as additional parameter in the system.                      
                                                                              
Using the steady state condition $\dot{y}=-vy'$, the shape equation reads     
\begin{equation}                                                              
\kappa=\frac{\sigma_{ik}u_{ik}}{2\gamma} + \frac{vy'}{D\left(                 
1+{y'}^2\right)^{1/2}},                                                       
\end{equation}                                                                
which is a nonlocal equation due to the long range elastic interactions.      
The boundary conditions at the tip are given by the arbitrary choice of the   
origin, $r(0)=r_0$, and $r'(0)=0$, since we are interested in symmetrical     
shapes, $r(\theta)=r(-\theta)$.                                               
Thus the entire shape is a function depending only on the parameter $v$.      
                                                                              
On the other hand, in the tail region, the elastic stresses have decayed, and 
the shape equation therefore becomes simply $-vy'=Dy''$.                      
Its general solution, $y(x)=A+B\exp(-vx/D)$, contains a growing exponential   
which is inconsistent with the boundary conditions of the straight crack.     
Therefore the solution must be arranged such that $B=0$.                      
Notice that in contrast to the surface diffusion process \cite{Brener03} a    
finite opening $A$ cannot be excluded since we do not have to obey mass       
conservation here.                                                            
At a first glance, suppression of the exponential seems to lead to a          
selection of the steady state velocity $v$ as only available parameter.       
However, from the rescaling behavior explored above it follows immediately    
that $B\sim 1/v$ (notice that $B$ has the dimension of a length), and         
therefore a finite velocity cannot be selected.                               
Consequently, a steady state solution for a growing crack in the framework of 
the static theory of elasticity does not exist.                               
                                                                              
The situation is different for the dynamical theory of elasticity:            
Here velocity enters into the equation of motion not only as $vr_0/D$ but     
also as $v/v_R$.                                                              
Hence a second independent parameter exists to guarantee $B=0$.               
Since now a rescaling is impossible, both the propagation velocity $v$ and    
the tip radius $r_0$ are selected.                                            
                                                                              
A tip splitting is possible for high applied tensions due to a secondary ATG  
instability:                                                                  
Since $\sigma\sim Kr_0^{-1/2}$ in the tip region and the local ATG length is  
$L_G\sim E\gamma/\sigma^2$, an instability can occur, provided that the tip   
radius becomes of the order of the ATG length.                                
In dimensionless units, this leads to the prediction $\Delta_{split}\sim      
{\cal O}(1)$.                                                                 
                                                                              
We developed a phase field code together with elastodynamics to describe      
phase transformations under stress, including for example also martensitic
transformations.                       
In the limit of vanishing shear modulus in one of the phases, this approach   
can be used to study melting and solidification processes which are induced   
by elastic forces \cite{Kassner01}.                                           
For a very soft secondary phase, crack propagation can be studied in the      
framework of a continuum theory, since then the usual boundary conditions of  
vanishing normal and shear stress are recovered.                              
Let $\phi$ denote the phase field with values $\phi=0$ for the soft and       
$\phi=1$ for the hard phase.                                            
The energy density contributions are                                          
$f_{el}=\mu(\phi)u_{ij}^2+\lambda(\phi)(u_{ii})^2/2$                          
for the elastic energy, with $\mu(\phi)=h(\phi)\mu^{(1)} +                    
(1-h(\phi))\mu^{(2)}$ and $\lambda(\phi)=h(\phi)\lambda^{(1)} +               
(1-h(\phi))\lambda^{(2)}$, where $h(\phi)=\phi^2(3-2\phi)$ interpolates       
between the phases and the superscripts denote the bulk values.               
The surface energy is                                                         
$f_s(\phi)=3\gamma \epsilon (\nabla\phi)^2/2$                                 
with the interface width $\epsilon$.                                          
Finally,                                                                      
$f_{dw}=6\gamma\phi^2(1-\phi)^2/\epsilon$                                     
is the well-known double well potential.                                      
Thus the total potential energy is                                            
\begin{equation}                                                              
U = \int dV \left( f_{el}+f_s+f_{dw}\right).                                  
\end{equation}                                                                
The elastodynamic equations are derived from the energy by the variation with 
respect to the displacements $u_i$,                                           
\begin{equation}                                                              
\rho \ddot{u}_i=-\frac{\delta U}{\delta                                       
u_i},
\end{equation}                                                                
and the dissipative phase fields dynamics follows from                        
\begin{equation} \label{phase:eq1}                                            
\frac{\partial\phi}{\partial t} = -\frac{D}{3\gamma\epsilon} \frac{\delta     
U}{\delta\phi}.                                                               
\end{equation}                                                                
These equations lead in the limit $\epsilon\to 0$ to the correct sharp        
interface limit above.                                                        
For the case of static elasticity, this was carefully proved in               
\cite{Kassner01}.                                                             
                                                                              
For the numerical realization, we employ explicit representations of both the 
elastodynamic equations and the phase field dynamics.                         
The elastic displacements are defined on a staggered grid \cite{Virieux86}.   
The derivation of the elastodynamic equations of motion from a discretized    
action integral obeying invariance against time inversion guarantees long     
time stability.                                                               
                                                                              
We study crack growth in a rectangular geometry of a strip with fixed         
displacements at its upper and lower boundary.                                
Horizontally, the grid can be shifted in order to keep the crack tip always   
in the center of the system.                                                  
Thus, crack growth can be studied over long times in relatively small         
systems.                                                                      
Typical dimensions used here are $600\times 200$ grid points, the phase field 
interface width is $\epsilon=5\,\Delta x$ ($\Delta x$ is the lattice unit)    
and the Poisson ratio is $\nu=1/3$.                                           
The Rayleigh speed $v_R$ is normalized to one.                                
In the soft phase, we typically set the elastic constants to one percent of   
the values in the hard phase;                                                 
however, these values are qualitatively not significant.                      
Notice that after rescaling the equations of motion depend only on the        
driving force $\Delta$ and the kinetic coefficient $D$;                       
in the numerical realization, also the phase field width $\epsilon$ and the   
strip width $L$ appear.                                                       
                                                                              
\begin{figure}                                                                
\begin{center}                                                                
\epsfig{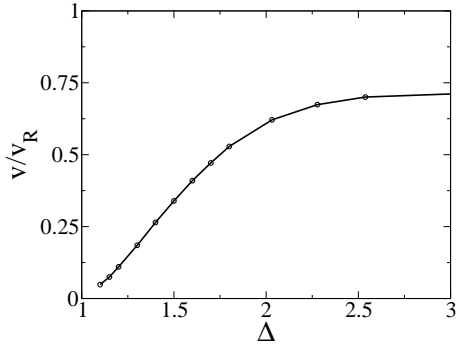}                      
\caption{Steady state velocity versus dimensionless driving force $\Delta$;   
$\Delta=1$ is the Griffith point. The tip radius is here determined by the    
phase field interface width. We used $D/\epsilon v_R=1.85$ here.              
For $\Delta\gtrapprox 3.4$ we get into the tip splitting regime.}             
\label{fig7}                                                                  
\end{center}                                                                  
\end{figure}

First, we studied the growth of cracks in the vicinity of the Griffith point. 
Here, the tip radius in not determined by the length scale $D/v_R$ but by the 
phase field interface width $\epsilon$.                                       
In the strip geometry, the dimensionless driving force is                     
$\Delta=u_0^2(\lambda+2\mu)/4L\gamma$ with the fixed vertical displacement    
$u_0$ applied to the strip.                                                   
As a nontrivial test, the numerical results validate the analytical           
prediction of the Griffith point $\Delta=1$ in the framework of the model, as 
the propagation velocity tends to zero (see Fig.\ \ref{fig7}).                
                                                                              
\begin{figure}                                                                
\begin{center}                                                                
\epsfig{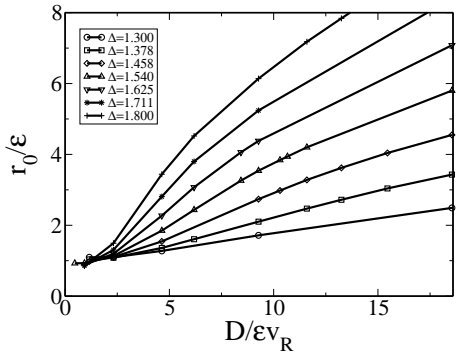}                                    
\caption{Tip radius $r_0$ as function of the kinetic coefficient $D$ for      
different driving forces $\Delta$. In the intermediate linear regime the      
length scales are well separated.}                                            
\label{fig3}                                                                  
\end{center}                                                                  
\end{figure}

\begin{figure}                                                                
\begin{center}                                                                
\epsfig{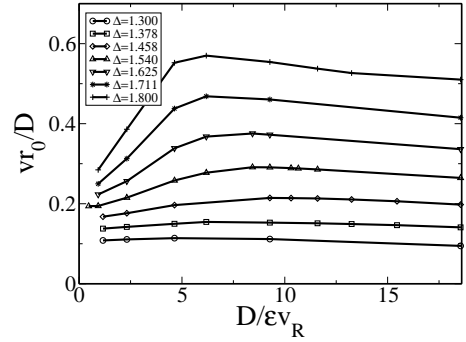}                                   
\caption{The quantity $vr_0/D$ as function of the kinetic coefficient for     
different driving forces $\Delta$.}                                           
\label{fig6}                                                                  
\end{center}                                                                  
\end{figure}

\begin{figure}                                                                
\begin{center}                                                                
\epsfig{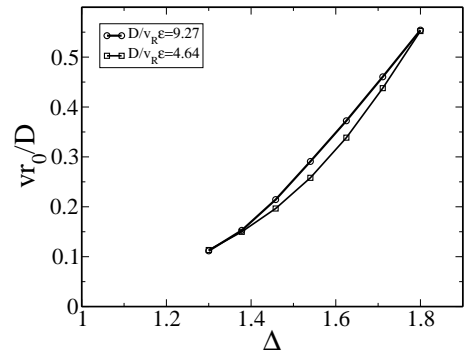}                                
\caption{$vr_0/D$ as function of the driving forces $\Delta$; the curves for  
different kinetic coefficients do not differ significantly, in agreement with 
the theoretical prediction.}                                                  
\label{fig8}                                                                  
\end{center}                                                                  
\end{figure}

The main goal was to approve that elastodynamics allows steady state growth   
without collapsing into the finite time cusp singularity of the ATG           
instability, selecting both a non-zero tip radius and a propagation velocity    
below the Rayleigh speed.                                                     
The simulations confirm this prediction, and a typical steady state shape is  
shown in Fig.\ \ref{fig2}.                                                    
Obviously, the tip radius is not determined by the intrinsic phase field      
length scale $\epsilon$.                                                      
This can also be seen in Fig.\ \ref{fig3}, where we plotted the steady state  
tip radius $r_0$ as function of the kinetic coefficient $D$ for various       
driving forces $\Delta$.                                                      
Only for very low kinetic coefficients, the tip radius is cut off by the      
interface width $\epsilon$, otherwise it is fairly bigger;                    
for high kinetic coefficients the saturation is induced by the system size.   
In between, however, the scales are well separated and the radius $r_0$ is a  
linear function of $D$, in good agreement with the theoretical analysis:      
since both parameters $v/v_R$ and $v r_0/D$ are predicted to be universal     
functions of the driving force $\Delta$ alone, we conclude that $r_0$ should  
depend linearly on the kinetic coefficient.                                   
Notice that also the tail opening $A$ is of 
the scale $A \sim D/v_R$ instead of      
$A \sim \epsilon$.                                                              
Furthermore, the ``dissipation rate'' $vr_0/D$ is rather independent of the   
kinetic coefficient (Fig.\ \ref{fig6}).                                       
Thus the universal dependence of $vr_0/D$ as a function of the driving force  
$\Delta$ can be extracted (Fig.\ \ref{fig8}).                                 
We believe that the results can be improved by increasing the system size and 
by a better separation of the length scales, which will be done in the near      
future.                                                                       
                                                                              
\begin{figure}                                                                
\begin{center}                                                                
\epsfig{file=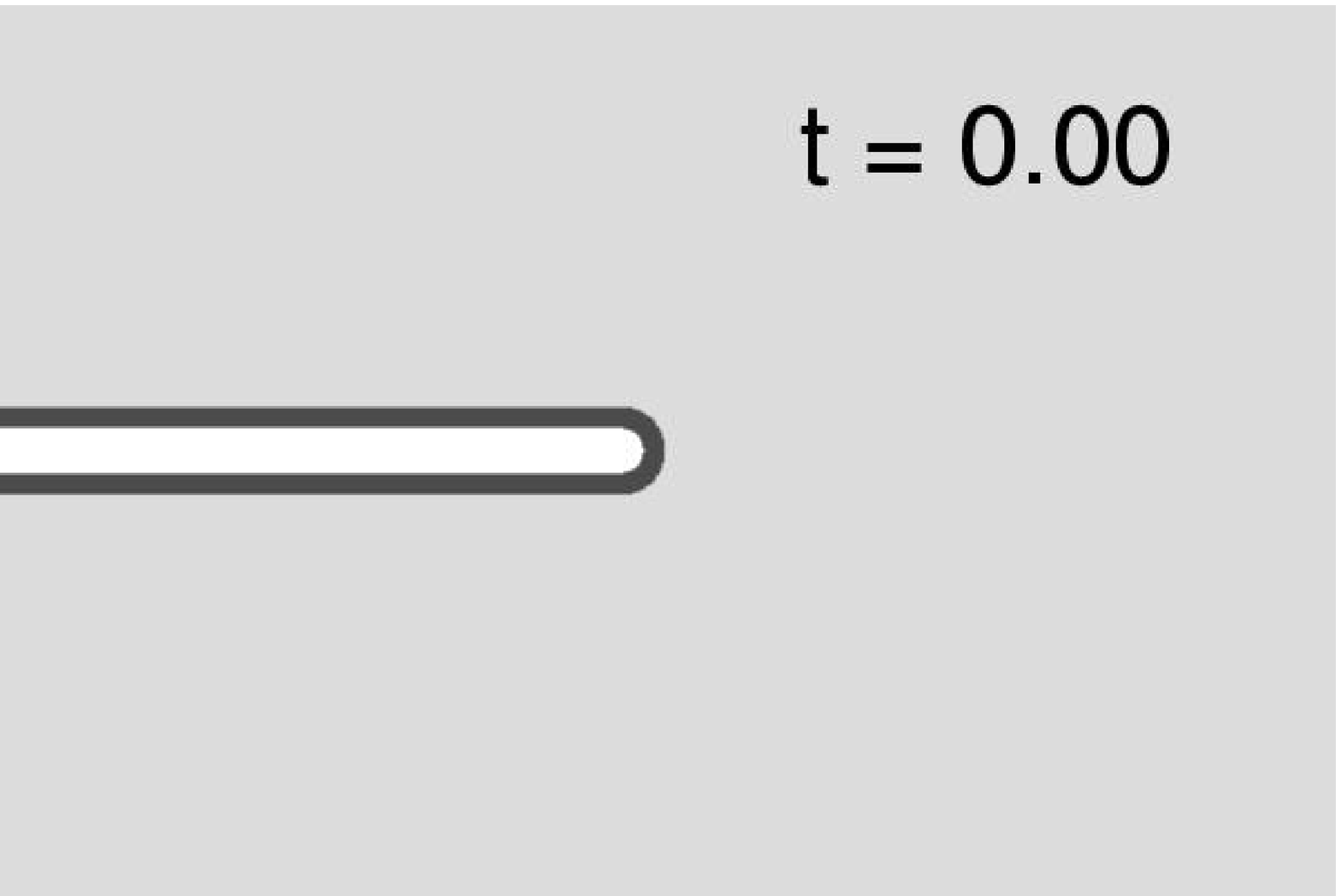, width=3.5cm}                                    
\epsfig{file=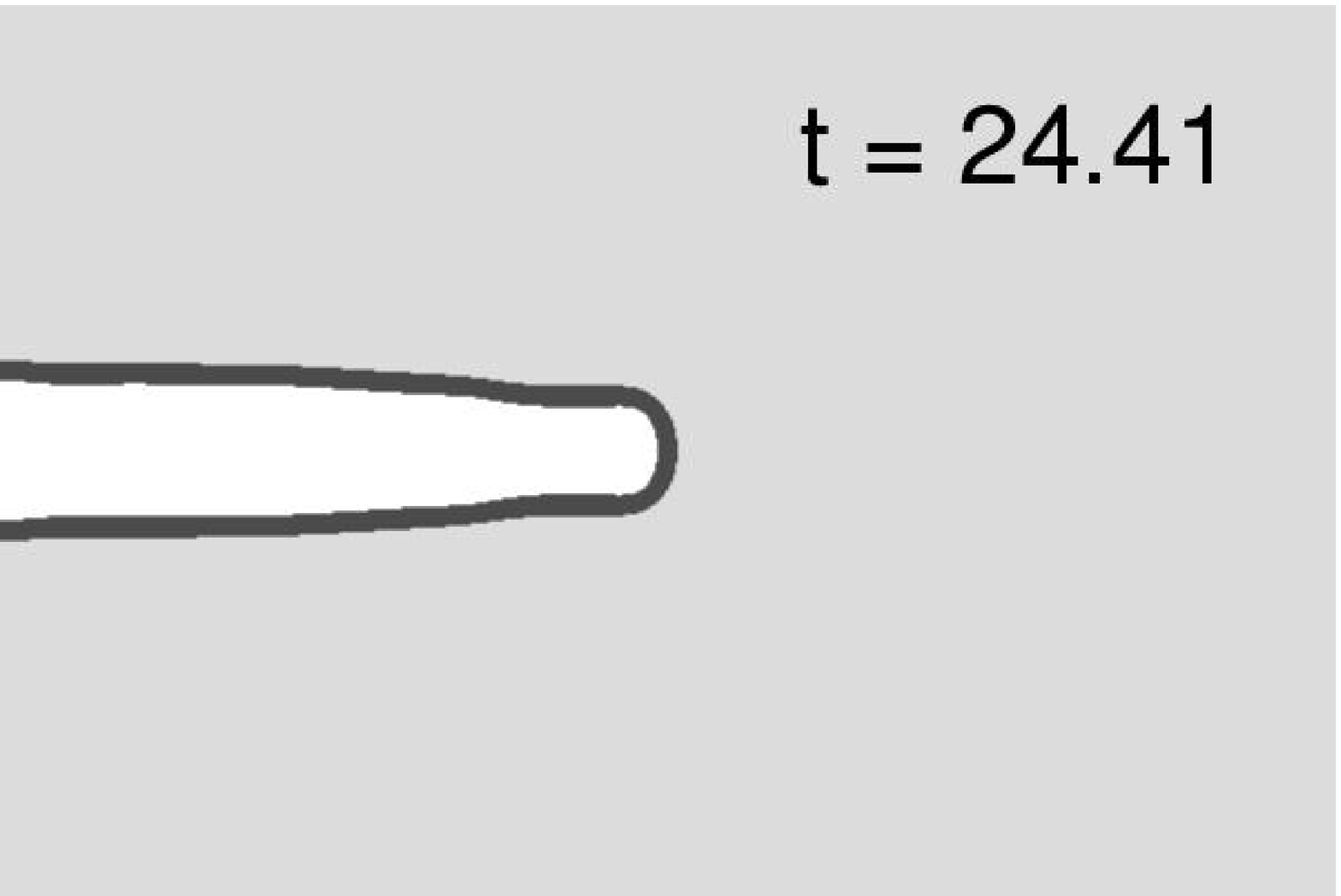, width=3.5cm}                                    
\epsfig{file=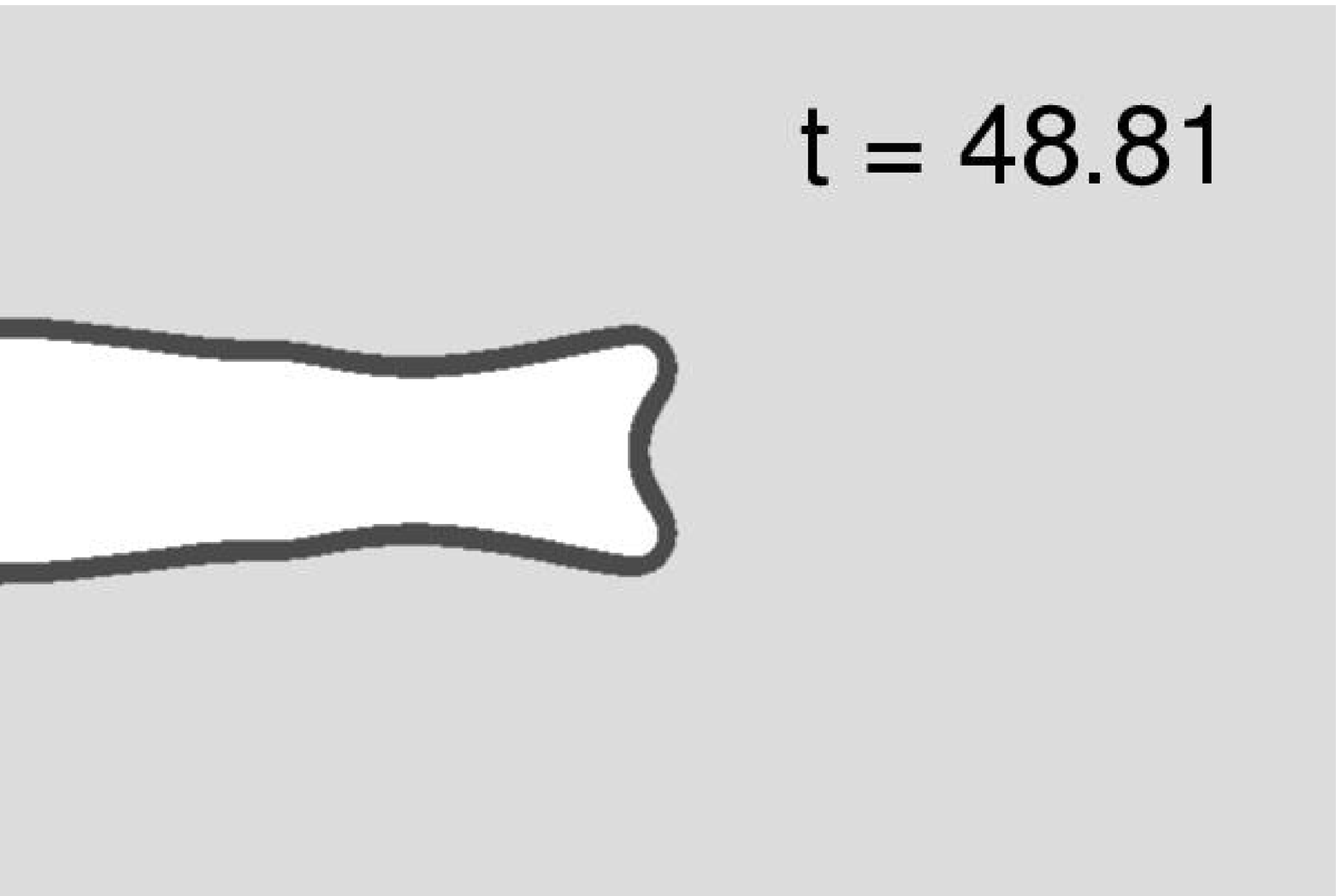, width=3.5cm}                                    
\epsfig{file=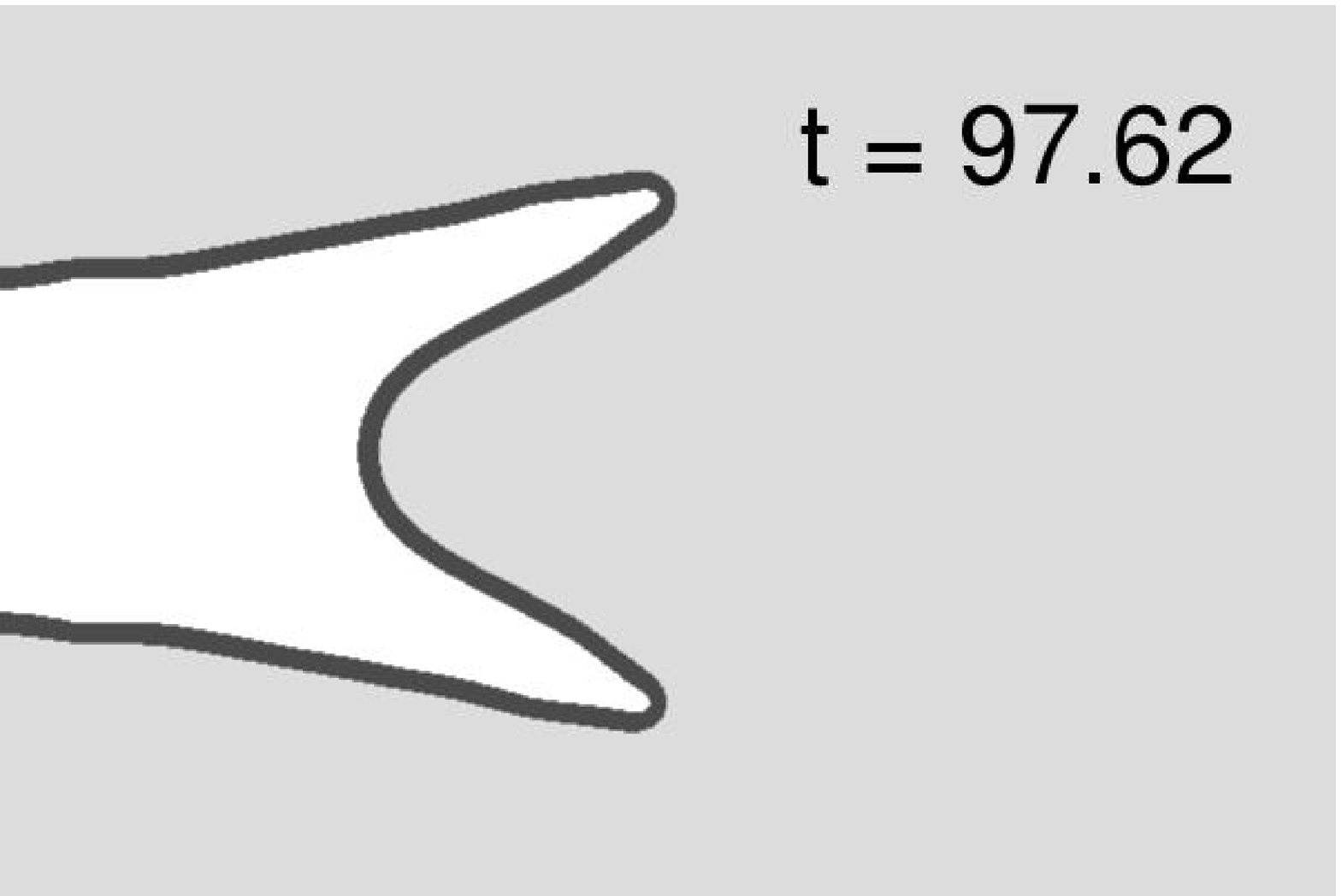, width=3.5cm}                                    
\epsfig{file=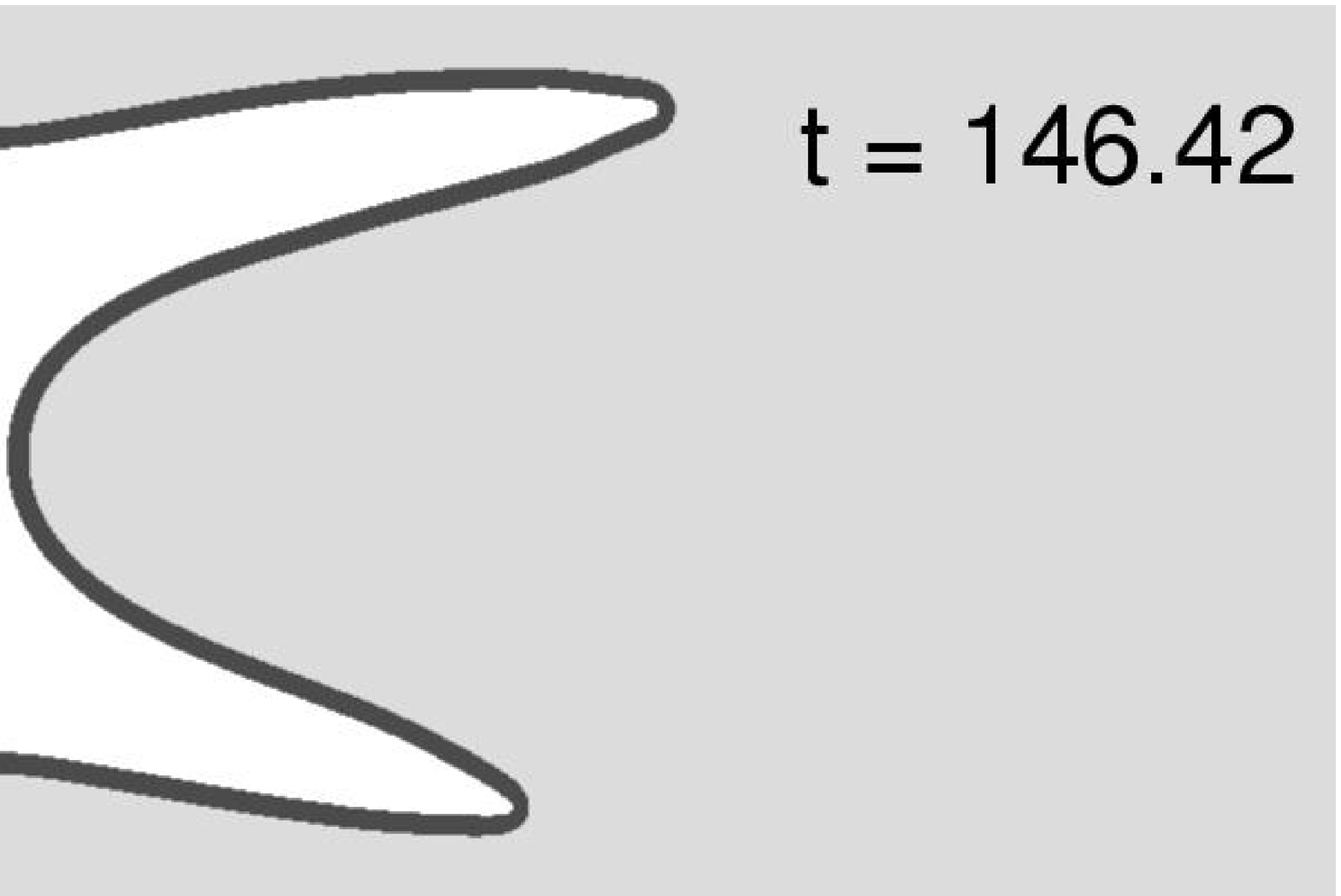, width=3.5cm}                                    
\epsfig{file=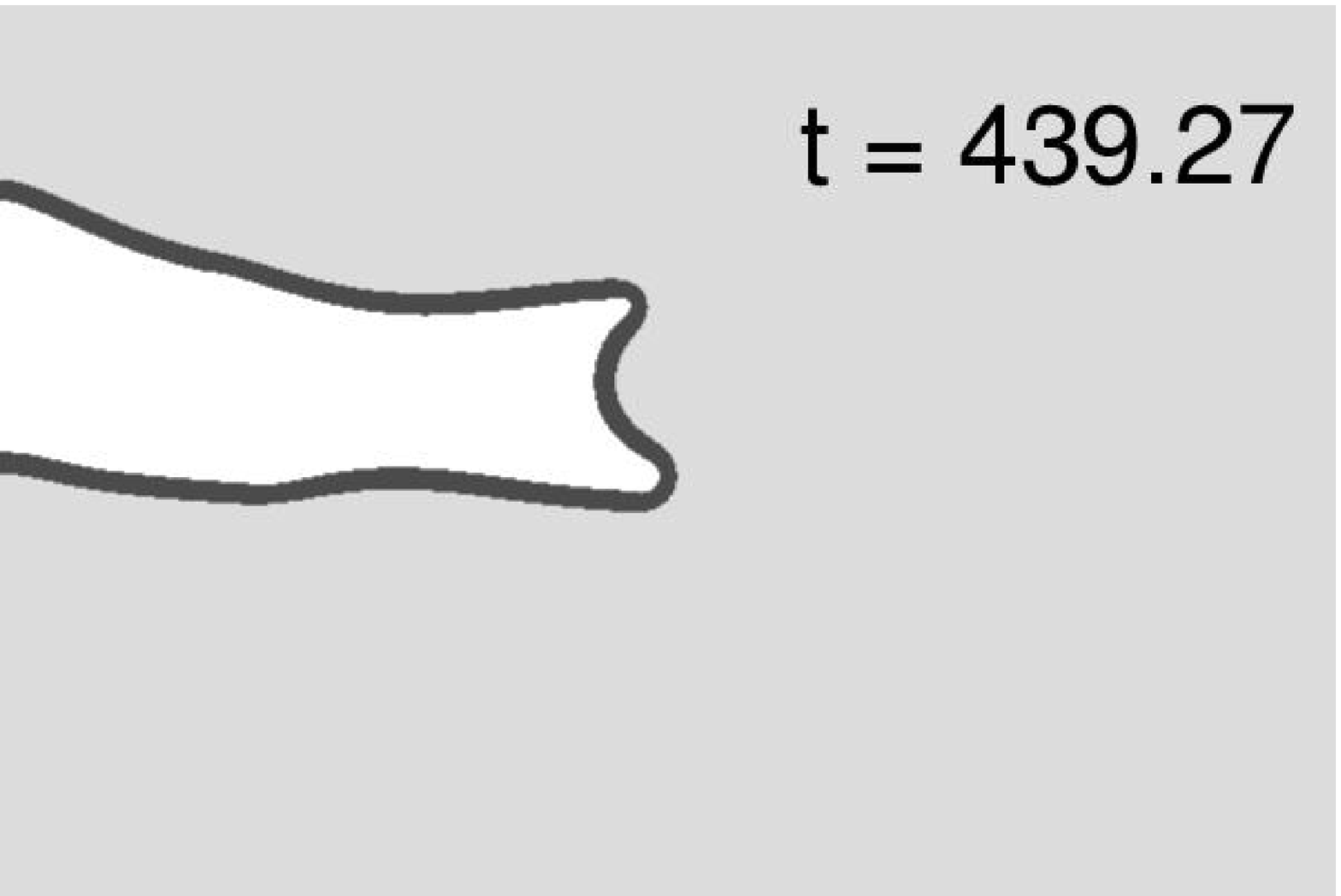, width=3.5cm}                                    
\caption{Irregular tip splitting scenario. We used $D/\epsilon v_R=1.85$ and  
$\Delta=3.6$; the system size is $600\times 400$. Time is given in units      
$D/v_R^2$.}                                                                   
\label{fig5}                                                                  
\end{center}                                                                  
\end{figure}

Snapshots of a typical tip splitting scenario for relatively high driving     
forces are shown in Fig.\ \ref{fig5}.                                         
Repeated irregular splitting of the crack tip occurs, followed by symmetrical 
growth of the sidebranches.                                                   
After a while, one finger wins the competition, moves back to the center of   
the strip and can finally split again.                                        
                                                                              
In summary, a phase field model has been developed to describe crack growth,  
based on thermodynamically well defined equations with a valid sharp          
interface limit. The model shows the possibility of 
steady state growth of cracks and a tip       
splitting instability, in agreement with analytical predictions.              
In contrast to other models previously discussed 
it provides a selection of the  
tip radius by the scale $D/v_R$.                                              
                                                                              
We thank Herv\'e Henry for useful discussions.                           
This work has been supported by the Deutsche Forschungsgemeinschaft under     
grant SPP 1120.                                                               
                                                                              


\begin{thebibliography}{9}                                                    
\bibitem{Griffith21}                                                          
A. A. Griffith, Philos. Trans. R. Soc. A, {\bf 21}, 163 (1921).               
\bibitem{Freund98}                                                            
L. B. Freund, {\em Dynamic Fracture Mechanics}, Cambridge University Press,   
1998.                                                                         
\bibitem{Hauch99}                                                             
J. Hauch et al., Phys.\ Rev.\ Lett.\ {\bf 82}, 3823 (1999).                   
\bibitem{Karma}                                                               
A. Karma, D. Kessler, and H. Levine, Phys.\ Rev.\ Lett.\ {\bf 87}, 045501     
(2001);                                                                       
A. Karma and A. Lobkovsky, Phys. Rev. Lett. {\bf 92}, 245510 (2004).          
\bibitem{Henry04}                                                             
H. Henry and H. Levine, Phys. Rev. Lett. {\bf 93}, 105504 (2004).             
\bibitem{Aranson00}                                                           
I. S. Aranson, V. A. Kalatsky, and V. M. Vinokur, Phys. Rev. Lett. {\bf 85},  
118 (2000);                                                                   
L. Eastgate et al., Phys. Rev. E {\bf 65}, 036117 (2002).                     
\bibitem{Kassner01}                                                           
K. Kassner et al., Phys. Rev E {\bf 63}, 036117 (2001).                       
\bibitem{Marconi05}                                                           
V. I. Marconi and E. A. Jagla, Phys. Rev. E {\bf 71}, 036110 (2005).          
\bibitem{Fineberg99}                                                          
J. Fineberg and M. Marder, Phys.\ Rep.\ {\bf 313}, 1 (1999).                  
\bibitem{Brener03}                                                            
E. A. Brener and R. Spatschek, Phys. Rev. E {\bf 67}, 016112 (2003).          
\bibitem{ATG}                                                                 
R. J. Asaro and W. A. Tiller, Metall. Tran. {\bf 3}, 1789 (1972);             
M. A. Grinfeld, Sov. Phys. Dokl. {\bf 31}, 831 (1986).                        
\bibitem{Nozieres93}                                                          
P. Nozi\`eres, J. Phys. I France {\bf 3}, 681 (1993).                         
\bibitem{Yang}                                                                
W. H. Yang and D. J. Srolovitz, Phys. Rev. Lett. {\bf 71}, 1593 (1993);       
B. J. Spencer and D. I. Meiron, Acta Metall. {\bf 42}, 3629 (1994);           
K. Kassner and C. Misbah, Europhys. Lett. {\bf 28}, 245 (1994).               
\bibitem{Virieux86}                                                           
J. Virieux, Geophys. {\bf 51}, 889 (1986).                                    
\end{thebibliography}
\end{document}